\documentclass[twocolumn,showpacs,preprintnumbers,amsmath,amssymb,aps]{revtex4-1}
\usepackage{graphicx}
\usepackage{dcolumn}
\usepackage{bm}
\usepackage[colorlinks=true]{hyperref}
\usepackage{epstopdf}
\usepackage[T1]{fontenc} 
\usepackage{lmodern}
\usepackage{color}

\begin{document}

\title{Impact of the impurity symmetry on orbital momentum relaxation and orbital Hall effect studied by the quantum Boltzmann equation.  }

\author{V.~V.~Kabanov}
\affiliation{Jozeh Stefan Institute, 1000 Ljubljana, Slovenia}
\author{A.~V.~Shumilin}
\affiliation{Jozeh Stefan Institute, 1000 Ljubljana, Slovenia}
\email[Electronic address: ]{andrei.shumilin@ijs.si}

\begin{abstract}
We develop a quantum Boltzman equation approach that incorporates microscopic impurity models into the theory of orbital transport,revealing, how impurity properties, including their symmetry, influence the relaxation of orbital momentum and the orbital Hall effect. Specifically, we demonstrate that when the impurity potential has axial symmetry, the relaxation is governed by the Dyakonov-Perel mechanism. In contrast,  when this symmetry is broken, scattering can result in a rapid Elliot-Yafet relaxation of orbital momentum. The details of
impurity potential also affect the intrinsic orbital Hall effect even when the impurity concentration is very small. Impurities that alter the orbital texture can also give rise to a skew-scattering contribution to the orbital Hall effect, although this does not necessarily dominate in materials that are nearly pristine due to the Dyakonov-Perel relaxation.
\end{abstract}

\maketitle
\section{Introduction}
\label{sec:intro}

Orbitronics is an emerging field within solid-state physics that studies the transport of the orbital angular momentum of electrons \cite{Orbitron1,Orbitron2}. It complements  the more established areas of electronics and spintronics, which study the transport of electron charge and spin respectively. Despite being relatively new - originating in the early 2000-th \cite{p-Si} - orbitronics holds significant promise for enhancing the  performance of devices that traditionally rely on spintronics, particularly those utilizing spin-orbit interaction. These devices include both the relatively developed technologies like magnetic random access memory (MRAM)\cite{MRAM} and more advanced concepts such as spin-orbit torque logic and artificial synapses \cite{SOTdev, SOTdev2}.

Like spin, orbital momentum is a form of angular momentum, and the physical phenomena involving it have a phenomenology similar to the spin-related phenomena. For instance, the recently reported orbital Hall effect \cite{OHALL-Sala, OtoTi, OHall-Zr} and orbital inverse Rashba-Edelstien effect \cite{OiRaEd} are analogous to the spin Hall effect and the conventional inverse Rashba-Edelstein effect, respectively.
The proposed orbital Rashba effect \cite{ORA11,ORA24} mirrors the conventional spin Rashba effect, occurring when inversion symmetry is broken, either in the bulk or at surfaces.

One notable advantage of orbital momentum over spin is that its manipulation does not necessarily require spin-orbit interaction, which is a relativistic effect \cite{Spintronics} and is generally weak, even in materials like heavy metals (e.g., Pt, Ta) that are commonly used in spintronic applications. In contrast, manipulating orbital momentum can occur without the spin-orbit interaction, although in some cases, spin-orbit coupling can convert orbital momentum into spin \cite{Lee2021}. Despite its nascent stage, orbitronics has already demonstrated significant potential, with devices based on orbital transport producing strong torques in ferromagnetic materials \cite{OtoCu, OtoG, OTorqueCoNb, OtoMu}. Remarkably, these torques can be induced by light metals, such as Nb and Ti, which usually exhibit weak spin-orbit interaction and are not typically effective in spintronic devices.

However, unlike spin, orbital momentum is quenched in most of the materials, meaning that its expectation value is zero in all electron eigenstates, making orbital transport a quantum phenomena. Any theory describing orbital transport must therefore account for the  coherent combinations of electron states across different bands.
Currently,  tight-binding models are the most commonly used theoretical tools for studying orbital transport \cite{Go1,Go2,OHallBi}. Although successful, these models are generally limited to pristine materials and are thus best suited for studying only certain aspects of orbital momentum phenomena. In comparing orbital and spin transport, it's noteworthy that spin accumulation at interfaces, due to the spin Hall effect, is controlled by the interplay between the spin Hall effect and spin relaxation. Of the three conventional mechanisms of the spin Hall effect (intrinsic, skew-scattering, and side-jump), only the intrinsic mechanism is a material property, while the other two require electron scattering for their description [18]. Spin relaxation, too, is closely tied to scattering.

Probably the most conventional theoretical tool for describing electron scattering and its effect on transport is the Boltzmann kinetic equation. However, the conventional Boltzmann equation only applies to electrons in their eigenstates, making it unsuitable for the orbital transport. A generalization of the kinetic equation that includes the coherent combinations of electron states across different bands is known as the quantum Boltsmann equation.  Although this approach has been less popular, it is known since the 1950s  \cite{Silin,Transport}. While being more complicated than the conventional Boltzmann equation, it still benefits from the quasi-classical approximation and is simpler than other techniques, such as the Green function formalism, which are sometimes used to describe impurity effects on transport involving several bands \cite{Tanaka, Nava}.

The quantum kinetic equation has been applied to orbital transport, albeit with a simplified phenomenological model for orbital momentum relaxation  \cite{Okin22}. A drawback of this approach approach is its inability to relate microscopic impurity structure to relaxation properties. This article aims to address this gap. We begin by formulating several microscopic impurity models within the tight-binding approach. We then derive the scattering operator based on these models and examine how impurity properties influence orbital momentum relaxation and the orbital Hall effect. Our findings show that impurity symmetry significantly affects orbital momentum relaxation. For example, when an impurity has axial symmetry, orbital momentum relaxation follows the Dyakonov-Perel mechanism, consistent with recent predictions \cite{orbDP}. However, when this symmetry is broken, the Elliot-Yafet relaxation mechanism can emerge, sometimes leading to a complete loss of orbital polarization in a single scattering event. Additionally, impurities that modify so-called orbital texturing \cite{Go1,orbTex} can give rise to the skew-scattering orbital Hall mechanism. Similar to the skew-scattering spin Hall effect, this mechanism requires consideration of scattering beyond the Born approximation.

The article is organized as follows: Sec.~\ref{sect:model} describes the tight-binding model, including the impurity potentials relevant to our discussion. Sec.~\ref{sect:EY} simplifies the model for the specific case of small $p$-band splitting, providing a toy model to analytically demonstrate how the Elliot-Yafet orbital momentum relaxation arises from impurity asymmetry. Sec.~\ref{sect:skew} modifies the toy model to incorporate the skew-scattering orbital Hall effect. In Sec.~\ref{sect:gen} we apply numerical simulations to the general model introduced in Sec.~\ref{sect:model} to explore how orbital momentum relaxation and the skew-scattering orbital Hall effect are affected outside the approximations made in Sec.~\ref{sect:EY} and \ref{sect:skew}. Sec.~\ref{sect:dis} presents a general discussion of our results.

\section{Tight-binding model and quantum kinetic equation}
\label{sect:model}

\begin{figure}[t]
  \centering
  \includegraphics[width=0.45\textwidth]{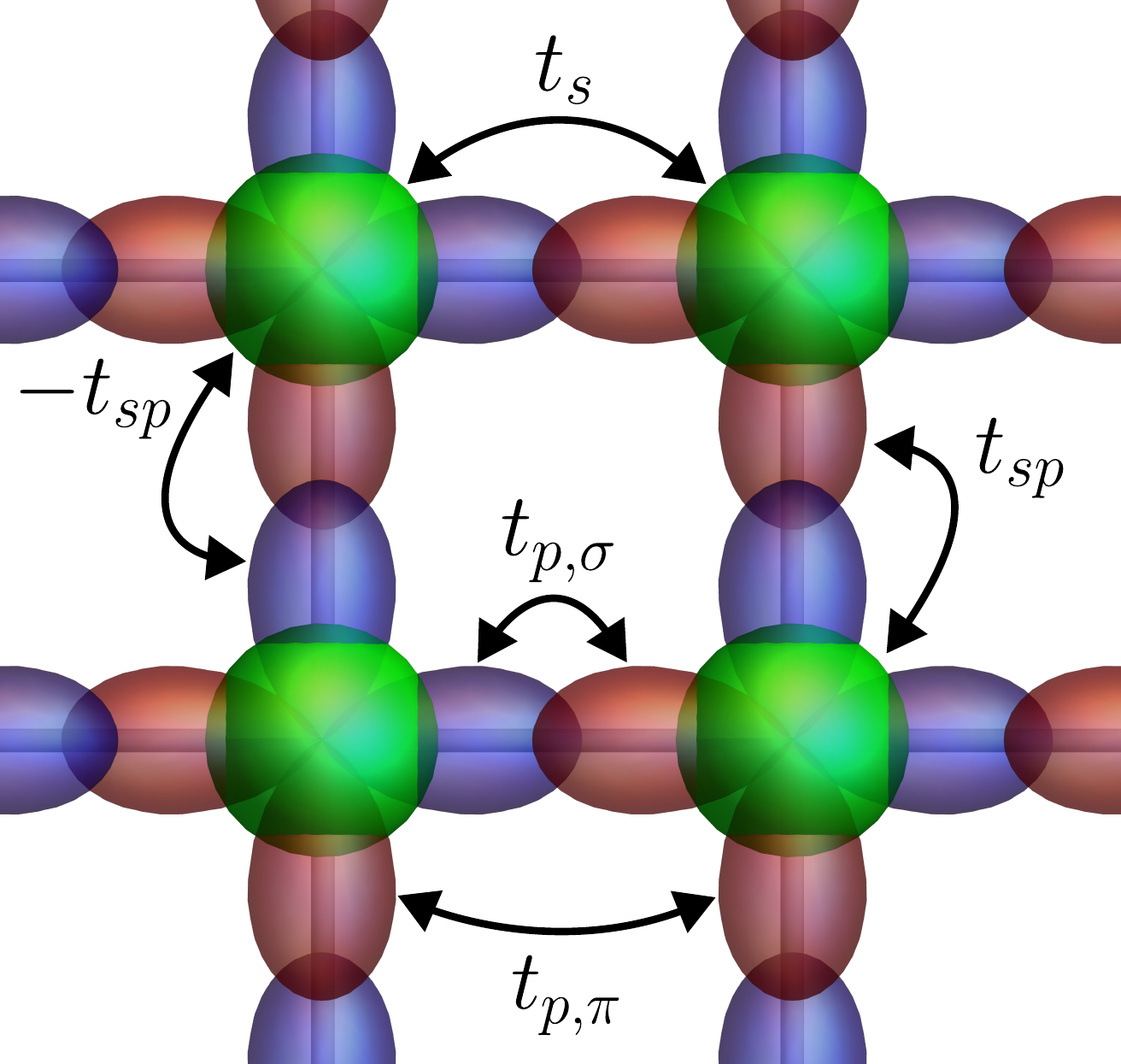}
  \caption{The tight-binding model described by Eqs.~(\ref{H}-\ref{Hy}). $s$ and $p$ - states are shown with green and blue/red colors respectively. The hopping integrals are shown with arrows.}
  \label{fig:mod}
\end{figure}

In this article, we focus on the minimal 2D tight-binding model that incorporates orbital texturing and, therefore, includes the orbital Hall effect. This model is similar to the one proposed in \cite{Go} and is depicted in Fig.~\ref{fig:mod}.  We consider a square lattice in the $xy$-plane composed of atoms with three important atomic orbitals. The $s$-orbital with orbital momentum $L=0$ has on-site energy $E_s$,  while  $p_x$ and $p_y$ orbitals correspond to $L=1$ have on-site energy $E_p$. In our model, only $z$-component $L_z$ of the orbital angular momentum can be accumulated or transported. This component is related to the coherent quantum superpositions of $p_x$ and $p_y$ states.

In addition to the energies $E_s$ and $E_p$, the model includes hopping integrals between neighboring $s$-orbitals ($t_s$) and between $p$-orbitals ($t_{p \sigma}$ and $t_{p\pi}$). Here $t_{p\sigma}$ corresponds to the $\sigma$-bonding of $p$-orbitals and $t_{p\pi}$ corresponds to $\pi$-bonding. We also include the hopping integral between $s$ and $p$ states ($t_{sp}$). Due to this hoping the band structure of the Hamiltonian eigenstates depend on the electron wavevector which is called orbital texturing. This texturing is crucial for the orbital Hall effect. Note that due to the symmetry of the $p$-orbitals, the sign of this integral is opposite for opposite hopping directions, as illustrated in Fig.~\ref{fig:mod}.

\begin{figure*}
  \centering
  \includegraphics[width=0.98\textwidth]{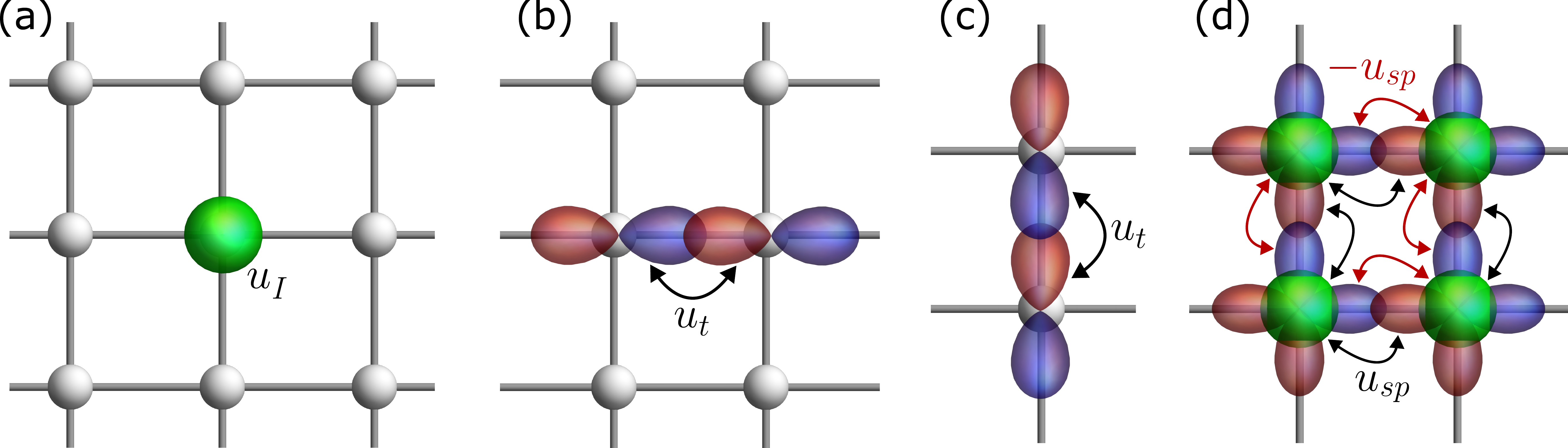}
  \caption{The models of the impurity potential. (a) The symmetric impurity with modified $E_s$ and $E_p$. (b, c) the asymmetric impurity corresponding to the modification of $t_{p,\sigma}$ in x-direction (b) or in y-direction (c). (d) the orbital texture impurity with 8 hopping integrals between $s$ and $p$ states modified by the value $u_{sp}$ or $-u_{sp}$ as shown by the arrows.}
  \label{fig:imp}
\end{figure*}

The Hamiltonian of an electron with wavevector ${\bf k}$ in our model is given by the  expression
\begin{equation}\label{H}
H = H_0 + H_x(k_x) + H_y(k_y)
\end{equation}
Here $k_x$ and $k_y$ are the wavevector components along $x$ and $y$ coordinates respectively. $H_0$ is the on-site part of the Hamiltonian. In the basis $(s, p_x, p_y)$ it equals
\begin{equation}\label{H0}
H_0 = \left(
\begin{array}{ccc}
E_s & 0 & 0 \\
0 & E_p & 0 \\
0 & 0 & E_p
\end{array}
\right),
\end{equation}
The term $H_x$ describes the electron movement along $x$-axis
\begin{multline}\label{Hx}
H_x = 2\cos(k_xa)\left(
\begin{array}{ccc}
t_s  & 0  & 0 \\
0 & t_{p,\sigma}  & 0 \\
0 & 0 & t_{p,\pi}
\end{array}
\right) \\ + 2 \sin(k_xa) t_{sp}
\left(
\begin{array}{ccc}
0  & i  & 0 \\
-i & 0  & 0 \\
0 & 0 & 0
\end{array}
\right)
\end{multline}
Here $a$ is the lattice constant. $H_y$ is the similar term describing the electron movement in $y$-direction.
\begin{multline}\label{Hy}
H_y = 2\cos(k_y a)\left(
\begin{array}{ccc}
t_s  & 0  & 0 \\
0 & t_{p,\pi}  & 0 \\
0 & 0 & t_{p,\sigma}
\end{array}
\right) \\ + 2 \sin(k_y a) t_{sp}
\left(
\begin{array}{ccc}
0  & 0  & i \\
0 & 0 & 0 \\
-i & 0 & 0
\end{array}
\right)
\end{multline}

The quantum Boltzmann equation is a generalization of the kinetic equation that allows for the treatment of coherent quantum superpositions of electron states while still utilizing the quasi-classical approximation -- i.e., assuming that the variation of the electron distribution is slow in space and time \cite{Silin,Transport}. In this article, we focus on the homogenous problems where neither the Hamiltonian nor the distribution function depends on coordinates. In this case, the quantum Boltzmann equation for a material subjected to a weak electric field ${\bf E}$ reads
\begin{equation}\label{kin1}
\frac{\partial f_1}{\partial t}  + \frac{e {\bf E}  }{\hbar} \frac{\partial f_0}{\partial {\bf k}} + \frac{i}{\hbar}[H, f_1]  = \hat{I} f_1,
\end{equation}
where $f_0$ is the equilibrium distribution function
\begin{equation}\label{f0}
f_0 = \left( 1 + \exp(H/T) \right)^{-1}.
\end{equation}
Here $f_1$ is the perturbation of the distribution function caused by the applied electric field ${\bf E}$ or due to non-equilibrium initial conditions. Both $f_0$ and $f_1$ depend on the wavevector ${\bf k}$ and are considered as $3\times 3$ matrices in the space of $s$, $p_x$ and $p_y$ bands. Thus, they encapsulate information about the quantum superpositions of electron states across different bands but are diagonal over the wavevector, which is the key aspect of the quasi-classical approximation. $\hat{I}$ denotes the scattering operator, which is described below.

The orbital momentum carried  by the electrons with wavevector ${\bf k}$ is
\begin{equation}\label{Lz}
\langle L_z(k) \rangle = {\rm Tr} f_1({\bf k}) \widehat{L}_z, \quad \widehat{L}_z = \hbar
\left(
\begin{array}{ccc}
0  & 0  & 0 \\
0 & 0  & -i \\
0 & i & 0
\end{array}
\right)
\end{equation}
It equals zero for all the eigenstates of the Hamiltonian $H$, and, therefore, $L_z=0$ when the distribution function is a diagonal matrix. Because $f_0$ is diagonal, there is no orbital polarization in the equilibrium.
However, due to the orbital texturing, the derivative $\partial f_0/\partial {\bf k}$ is not diagonal. It allows the generation of orbital polarization by the applied electric field.

 To derive the general expression for the scattering operator $\hat{I}$ we follow the procedure introduced in \cite{Mineev,MineevJETP} for describing  spin transport in metals without inversion symmetry. The detailed derivation is provided in Appendix \ref{app:I3}. Within the Born approximation, it yields.
\begin{multline}\label{Scat1}
  \left( \hat{I} f \right)_{\alpha\beta}({\bf k}) = \frac{\pi}{\hbar} N_I \sum_{{\bf k}',\xi,\zeta} V_{\alpha\xi}({\bf k},{\bf k}')f_{\xi\zeta}({\bf k}') V_{\zeta\beta}({\bf k}'{\bf k}) \\ \times \left[ \delta\bigl(\varepsilon_\xi({\bf k}') - \varepsilon_\beta({\bf k})\bigr)
  + \delta\bigl(\varepsilon_\alpha({\bf k}) - \varepsilon_\zeta({\bf k}')\bigr)  \right] \\
  - V_{\alpha\xi}({\bf k},{\bf k}')V_{\xi\zeta}({\bf k}',{\bf k}) f_{\zeta\beta}({\bf k}) \delta\bigl( \varepsilon_\xi({\bf k}') - \varepsilon_\beta({\bf k}) \bigr) \\ -
  f_{\alpha\xi}({\bf k}) V_{\xi\zeta}({\bf k},{\bf k}') V_{\zeta\beta}({\bf k}',{\bf k}) \delta\bigl(\varepsilon_\alpha({\bf k}) - \varepsilon_\zeta({\bf k}')\bigr)
\end{multline}
Here $N_I$ is the total number of impurities. The Greek letters $\alpha, \,\beta\, ...$ denote the eigenstates of the Hamiltonian for a given wavevector. $\varepsilon_\alpha ({\bf k})$ is the energy of the $\alpha$-th eigenstate of the Hamiltonian corresponding to the wavevector ${\bf k}$. $V_{\alpha\xi}({\bf k},{\bf k}')$ is the matrix element of the impurity potential corresponding the the $\alpha$-th state of the Hamiltonian with momentum ${\bf k}$ and the $\xi$-th state of the Hamiltonian with the momentum ${\bf k}'$. Note that the bases of the eigenstates for the wavevectors ${\bf k}$ and ${\bf k}'$ are different.

\subsection{Impurity models}

One of the goals of this article is to relate the studied physical phenomena to the microscopic properties of the impurities, including their symmetries. To achieve this, we formulate several models of the impurity potentials within the tight-binding approach and incorporate them into the quantum Boltzmann equation through the scattering operator $\hat{I}$. These models are illustrated in Fig.~\ref{fig:imp}. In contrast to Eq.~(\ref{Scat1}), in this section, we provide the matrix elements of the impurity potentials in the ''laboratory'' basis of $s$, $p_x$ and $p_y$ states. The transition between these bases should be performed where necessary.

The first and simplest impurity potential considered is the "symmetric impurity". This potential represents an additional potential energy $u_I$ on a single site of the tight-binding lattice that does not depend on band. Such an impurity has axial symmetry. The corresponding  matrix elements are expressed as follows
\begin{equation}\label{V0}
V^{(sym)}({\bf k},{\bf k}') = u_I\frac{a^2}{{\cal S}} \hat{1}.
\end{equation}
Here ${\cal S}$ is the sample area.  $\hat{1}$  is the unit matrix in the band space. Note that $V^{(sym)}({\bf k},{\bf k}')$ commutes with $L_z$.

The "asymmetric impurity" is the second impurity model, representing a modified $t_{p,\sigma}$ bond in the tight-binding model. This impurity does not have axial symmetry and instead favors either $x$ or $y$ axis.
When the modified bond is along $x$-axis, the impurity matrix elements are:
\begin{equation}\label{Vpx}
{V}^{(tx)}({\bf k},{\bf k}') =  u_t \frac{2a^2}{{\cal S}} \cos\frac{(k_x  + k_x') a}{2}
\left(
\begin{array}{ccc}
0  & 0  & 0 \\
0 & 1  & 0 \\
0 & 0 & 0
\end{array}
\right)
\end{equation}
Here $u_{t}$ is the modification of $t_{p,\sigma}$. The matrix structure of ${V}^{(tx)}$ shows that the impurity potential acts only on $p_x$-states. ${V}^{(tx)}$  does not commute with $L_z$.

A similar impurity with a modified  $t_{p,\sigma}$ bond in $y$-direction can be described by the following matrix elements:
\begin{equation}\label{Vpy}
{V}^{(ty)}({\bf k},{\bf k}') =  u_t \frac{2a^2}{\cal{S}} \cos\frac{(k_y  + k_y')a}{2}
\left(
\begin{array}{ccc}
0  & 0  & 0 \\
0 & 0  & 0 \\
0 & 0 & 1
\end{array}
\right)
\end{equation}

The third model, the "orbital texture impurity", is designed to study how the impurity-induced orbital texture can lead to the skew scattering orbital Hall effect. This model consists of the four nodes of tight binding lattice, arranged in square, with modified $s-p$ bonds. Its structure is shown in Fig.~\ref{fig:imp}(d). The expression for the matrix elements within this model is:
\begin{multline}\label{Vsp}
V^{(sp)}({\bf k},{\bf k}') =  u_{sp}\frac{4 a^2}{{\cal S}} \\ \times \left[ \sin\frac{k_xa + k_x'a}{2}\cos\frac{k_ya - k_y'a}{2}
\left(\begin{array}{ccc}
0 & i&0 \\
-i & 0 & 0 \\
0 & 0 & 0
\end{array}
\right) \right. \\
\left. +
\sin\frac{k_ya + k_y'a}{2}\cos\frac{k_xa - k_x'a}{2}
\left(\begin{array}{ccc}
0 & 0& i \\
0 & 0 & 0 \\
-i & 0 & 0
\end{array}
\right)
\right]
\end{multline}

\section{Elliot-Yafet orbital momentum relaxation}
\label{sect:EY}

The goal of this section is to comprehensively demonstrate the significance of the impurity symmetry in orbital momentum relaxation. To achieve this, we introduce a limiting case that further simplifies the model presented in Sec.~\ref{sect:model} making it amenable to analytical treatment.
We  assume that $t_{sp}  \ll |t_{p,\sigma} - t_{p,\pi}| \ll t_{p,\sigma}, t_{p,\pi}, t_{s}, |E_{s}-E_p|$. This assumption implies that the $p_x$ and $p_y$ states are nearly degenerate, while the $s$-states are significantly split from them by a large energy difference, $E_s-E_p$. Nonetheless, the $sp$-hopping term $t_{sp}$ in the Hamiltonian is even smaller than the difference between $t_{p,\sigma}$ and $t_{p,\pi}$, which ensures that the $p_x$ and $p_y$ states remain well-defined.

We also assume that the Fermi surface lies within the $p$-band and that the Fermi momentum is small enough to allow us to expand the sine and cosine functions in Eqs.~(\ref{Hx}, \ref{Hy}):
\begin{equation}\label{sc-exp}
\sin(k_{x,y}a) \approx k_{x,y}a, \quad \cos(k_{x,y}a) \approx 1 - \frac{(k_{x,y}a)^2}{2}.
\end{equation}

In this scenario, the $p_x$ and $p_y$ states play a decisive role in both charge transport and  orbital momentum transport. Accordingly, we reduce our Hamiltonian to these state by applying the Schrieffer - Wolff transformation $H' = \Pi_{p} e^{S_{SW}} H e^{-S_{SW}}$, where $\Pi_p$ is the projection operator onto the $p_x$ and $p_y$-states (which correspond to the second and third rows and columns of the Hamiltonian~(\ref{H}-\ref{Hy})) and $S_{SW}$ is the matrix:
\begin{equation}\label{trSF}
S_{SW} =  \frac{2iat_{sp}}{\Delta_{sp}({\bf k})}
\left(
\begin{array}{ccc}
0  & k_x  & k_y \\
k_x & 0  & 0 \\
k_y & 0 & 0
\end{array}
\right)
\end{equation}
Here $\Delta_{sp}({\bf k}) = \varepsilon_{s}({\bf k}) - \varepsilon_{px}({\bf k}) \approx \varepsilon_{s}({\bf k}) - \varepsilon_{py}({\bf k})$ is the difference between the energy $\varepsilon_s({\bf k}) = E_s + 2t_s\cos(k_x a) + 2t_s\cos(k_ya)$ of the $s$-state with a given momentum ${\bf k}$ and the energies $\varepsilon_{px}({\bf k})$ and  $\varepsilon_{py}({\bf k})$ of the $p_x$ and $p_y$ states. $\varepsilon_{px}({\bf k}) = E_p + 2t_{p,\sigma}\cos(k_x a) + 2t_{p,\pi}\cos(k_ya)$  and $\varepsilon_{py}({\bf k}) = E_p + 2 t_{p,\pi}\cos(k_x a) + 2 t_{p,\sigma}\cos(k_ya)$, however, we neglect their difference in Eq.~(\ref{trSF}).

The transformation (\ref{trSF}) yields the Hamiltonian of the $p$-states
\begin{equation}\label{Hp}
H_p({\bf k}) = \left(
\begin{array}{cc}
\varepsilon_{px}'({\bf k})  & -\lambda({\bf k})  \\
 -\lambda({\bf k}) & \varepsilon_{py}'({\bf k})
\end{array}
\right)
\end{equation}
Here
 $\lambda({\bf k}) = 2t_{sp}^2 k_xk_ya^2/\Delta_{sp}$, $\varepsilon_{px}' = \varepsilon_p^{(0)} - t_{p,\sigma}' k_x^2a^2 - t_{p,\pi} k_y^2a^2$ and
 $\varepsilon_{py}' = \varepsilon_p^{(0)} - t_{p,\pi}  k_x^2a^2 - t_{p,\sigma}' k_y^2a^2$ where we introduced the notations $\varepsilon_p^{(0)} = E_p + 2t_{p,\pi} + 2 t_{p,\sigma}$ and
$t_{p,\sigma}' = t_{p,\sigma} + 4t_{sp}^2/\Delta_{sp}$.
The Hamiltonian (\ref{Hp}) is written up to the terms proportional to $t_{sp}$ squared.

Within the second order approximation in $t_{sp}$ it can be diagonalized with a matrix $T({\bf k})$, $H_D({\bf k}) = T^+({\bf k})H_{p}({\bf k})T({\bf k})$, where
\begin{multline}\label{Hd}
H_D({\bf k}) = \left(
\begin{array}{cc}
\varepsilon_{px}'({\bf k})  & 0  \\
 0 & \varepsilon_{py}'({\bf k})
\end{array}
\right), \\
T({\bf k}) = \left(
\begin{array}{cc}
1  & \alpha({\bf k})  \\
 -\alpha({\bf k}) & 1
\end{array}
\right),
\end{multline}

Here $\alpha({\bf k}) = \lambda({\bf k})/(\varepsilon_{px}'({\bf k}) - \varepsilon_{py}'({\bf k}))$.
Eq.~(\ref{Hd}) represents the transition
to the eigenbasis of the Hamiltonian, which depends on ${\bf k}$. All the subsequent expressions in this section are presented in this basis.

After applying the Schrieffer - Wolff transformation, the operator $L_z = \hbar\sigma_y$ is proportional to the Pauli matrix $\sigma_y$. Since $T$ commutes with $\sigma_y$, the expression for $L_z$ remains unchanged across all eigenbases corresponding to different ${\bf k}$ values.

To explore the analytical solvability of this model, it is helpful to solve the quantum Boltzmann equation analytically under a static electric field,  with the scattering operator approximated by a single relaxation time $\hat{I}f \rightarrow -f/\tau$. This approximation does not mix different wavevectors ${\bf k}$ making the solution straightforward.
\begin{multline}\label{ftau}
f_1 = \left(
\begin{array}{cc}
g_x  & g_{(xy)}  \\
 g_{(xy)}^* & g_y
\end{array}
\right),  \\ g_{x,y}  = -e{\bf E}\tau {\bf v}_{x,y}  \frac{\partial f_0(\varepsilon_{px,py}')}{\partial \varepsilon_{px,py}'}
\end{multline}
Here ${\bf v}_{x,y} =  \hbar^{-1}\partial \varepsilon_{x,y}'/\partial {\bf k}$ is the velocity in $p_x$ and $p_y$ bands respectively, $f_0(x) = (1 + \exp(\frac{x-\mu}{T}))^{-1}$ is the Fermi function.
\begin{equation}\label{gxy}
g_{(xy)} =  \frac{e{\bf E}\tau (f_0(\varepsilon_{px}') - f_0 (\varepsilon_{py}'))  }{\hbar + i(\varepsilon_{px}' - \varepsilon_{py}')\tau} \frac{\partial \alpha}{\partial {\bf k}}
\end{equation}

Eqs.~(\ref{ftau},\ref{gxy}) illustrate the different roles of the diagonal components $g_x$, $g_y$ and the off-diagonal component $g_{(xy)}$ of the distribution function. At low temperatures, the Fermi function $f_0(x) \approx \theta(\mu-x)$ and its derivative $\partial f_0(x)/\partial x \approx -\delta(x-\mu)$, where $\theta$ is the Heaviside step function and $\delta$ is the Dirac delta function. As a result, the diagonal terms $g_x$ and $g_y$ are only significant  at the corresponding Fermi surfaces, representing electron and hole excitations in $p_x$ and $p_y$ bands, respectively. The off-diagonal part exists between the Fermi surfaces and represents the quantum coherent states of an electron. Consider a momentum ${\bf k}$ with $\varepsilon_{py}'({\bf k}) > \mu > \varepsilon_{px}'({\bf k})$. At zero temperature in equilibrium, the system would contain a single electron with momentum ${\bf k}$ in $px$ band. $g_{xy}$ corresponds to this electrons acquiring some amplitude to be in the $py$ band, effectively placing it in a coherent superposition of $px$ and $py$ states. This leads to  collective spin wave-like modes, sometimes called the Silin modes~\cite{SW12,SWjetp,SWgr}.
These modes are responsible for orbital transport, and the orbital momentum carried by electrons with momentum ${\bf k}$ is given by $-2\hbar{\rm Im}g_{xy}({\bf k})$.

The approximation  $|t_{p,\sigma} - t_{p,\pi}| \ll t_{p,\sigma}, t_{p,\pi}$  is used to confine $g_{(xy)}$ to a very narrow, albeit finite, strip in $k$-space between the Fermi surfaces of $p_x$ and $p_y$ bands. This allows us to neglect the dependence of the Hamiltonian, including its impurity part, on the absolute value of ${\bf k}$ in this region and introduce the distribution function ${G}(\varphi)$ integrated over $|{\bf k}|$:
\begin{equation}\label{G}
{G}(\varphi) =  \int f_1(k,\varphi) kdk
\end{equation}
With our approximations, it is possible to formulate a closed Boltzmann equation for $G(\varphi)$:
\begin{equation}\label{bolG}
\frac{\partial G}{\partial t} + \int \frac{e{\bf E}}{\hbar} \frac{\partial f_0}{\partial {\bf k}}kdk + \frac{i}{\hbar}[H_D{({\varphi})},G] = \hat{I}_GG
\end{equation}

The scattering operator $\hat{I}_G G$ for the integrated distribution function in the Born approximation is given by the expression
\begin{multline}\label{IG}
(\hat{I}_G G)(\varphi) = \frac{\gamma_p {\cal S} N_I}{ 2\hbar}\int_0^{2\pi} 2\widetilde{V}(\varphi\varphi')G(\varphi')\widetilde{V}(\varphi'\varphi) \\ - \widetilde{V}(\varphi\varphi')\widetilde{V}(\varphi'\varphi)G(\varphi)
-G(\varphi) \widetilde{V}(\varphi\varphi')\widetilde{V}(\varphi'\varphi) d\varphi',
\end{multline}
where $\widetilde{V}(\varphi\varphi')$ is the modified impurity scattering matrix element
\begin{equation}\label{Vtran}
\widetilde{V}(\varphi\varphi') = T^{+}({\varphi}) \Pi_p\left( e^{S_{SW}} V({\bf k},{\bf k}')e^{-S_{SW}} \right) T({\varphi}').
\end{equation}
Here $\varphi$ and $\varphi'$ are the polar angles corresponding to the wavevectors ${\bf k}$ and ${\bf k'}$ respectively.
$V({\bf k},{\bf k}')$ is assumed to be some combination of the potentials~(\ref{V0}-\ref{Vsp}).
We neglected the dependence of the Hamiltonian $H({\bf k})$, it's perturbation $V({\bf k},{\bf k}')$ and the transformation matrix $T({\bf k})$ on the absolute value of wavevectors. These
absolute values (for both ${\bf k}$ and ${\bf k}'$) are taken equal to the Fermi wavevector $k_p$ in $p$-bands. In particular it allows us to write $H_{D}(\varphi)$ and $T(\varphi)$ instead of $H_D({\bf k})$ and $T({\bf k})$ respectively. $\gamma_p(\varepsilon) = (2\pi)^{-1}\int \delta(\varepsilon_\alpha({\bf k})-\varepsilon)kdk$ is the density of the one branch of $p$-states. We neglect its dependence on $\alpha$, $\varepsilon$ and $\varphi$ and include into the equations only it's value at the Fermi level $\gamma_p \approx (4\pi t_{p,\sigma}a^2)^{-1} \approx (4\pi t_{p,\pi}a^2)^{-1}$.

Eq.~(\ref{IG}) has a significant advantage over Eq.~(\ref{Scat1}): it does not contain  delta functions that depend on the band indices. As a result, both the in-scattering and out-scattering terms can be expressed as matrix products, greatly simplifying the analysis of the interplay between the microscopic impurity properties and the orbital momentum relaxation.

The "symmetric" impurity model is transformed by the operation in Eq.~(\ref{Vtran}) into the matrix
\begin{equation}
\widetilde{V}^{(sym)}(\varphi,\varphi') = u_I T^{+}(\varphi)T(\varphi')
\end{equation}
Note that $\widetilde{V}^{(sym)}(\varphi,\varphi')\widetilde{V}^{(sym)}(\varphi',\varphi) = u_I^2$.  This implies that (I) the out-scattering process is independent of angle and band, and (II) the change in the distribution function due to scattering is reduced to its transformation into the new basis. Specifically, $L_z$ is conserved during scattering on symmetric impurities, but is redistributed across different polar angles $\varphi$.

When only symmetric impurities are present in the sample, the theory of orbital momentum relaxation can be mapped onto the theory of Dyakonov-Perel spin relaxation in solids with broken inversion symmetry \cite{DP0, DP1, Dyak}.
In this mapping, the orbital momentum corresponds to the $y$-direction of the spin, because $L_z$ is represented by $y$-Pauli matrix. It is conserved during the scattering, but between the scattering events, it precesses with a frequency $(\varepsilon_{px}' - \varepsilon_{py}')/\hbar$, which corresponds to an effective "magnetic field" directed along $z$-axis. $z$-component of spin in this mapping represents the different occupation probabilities of $px$ and $py$ bands and does not precess. $x$ component represents the $p$-states directed along arbitrary axis in $xy$-plane. However, the mapping is not perfect, as the $x$ and $z$-components of the effective spin are not conserved during the scattering due the fact that the $\sigma_x$ and $\sigma_z$ Pauli matrices do not commute with $T$.

Using this mapping, the relaxation of the orbital momentum is described by the equation
\begin{equation}\label{map}
\frac{\partial l(\varphi)}{\partial t} + i\omega(\varphi) l(\varphi) = \frac{\overline{l} - l(\varphi)}{\tau_{out}}
\end{equation}
Here $l(\varphi)$ is the "complex orbital momentum", its real and imaginary parts are defined
 as ${\rm Re}\,l(\varphi) =\hbar{\rm Tr}G(\varphi)\sigma_y $ and
${\rm Im}\,l(\varphi) =\hbar{\rm Tr}G(\varphi)\sigma_x$ respectively. $\overline{l} = (1/2\pi)\int_0^{2\pi}l(\varphi)d\varphi$ is the averaged value of $l(\varphi)$. $\omega(\varphi) = \omega_0\cos(2\varphi)$ is the angle-dependent precession frequency, where $\omega_0 = 2(t_{p,\sigma}' - t_{p,\pi})(1 - \cos(k_p a))/\hbar$. The orbital momentum $L_z$ equals to $L_z = 2\pi {\rm Re} \, \overline{l}$.
$\tau_{out}$ is the out-scattering time.

Eq.~(\ref{map}) can be easily solved numerically by expanding $l(\varphi)$ into Fourrier series. In particular when the out-scattering time is small we can keep only first two Fourrier components. In that case the relaxation of $L_z$ is described by the expression
\begin{equation}\label{map-DP}
L_z(t) = L_z(0){\rm Re} \left( \frac{\Omega_- e^{\Omega_+ t}  -  \Omega_+ e^{\Omega_- t}}{\Omega_- - \Omega_+}  \right),
\end{equation}
where $\Omega_\pm = (-1 \pm \sqrt{1 - 2\omega_0^2\tau^2_{out}})/2\tau_{out}$.

 The "asymmetric impurity" potential is
 is transformed by the operation in Eq.~(\ref{Vtran}) into
\begin{equation}\label{Vtxty}
\widetilde{V}^{(tx,ty)}(\varphi,\varphi') = 2u_t \frac{a^2}{{\cal S}} \xi^{(tx,ty)}_{\varphi,\varphi'} T^{+}(\varphi) \frac{1\pm\sigma_z}{2}T(\varphi')
\end{equation}
Here $\sigma_z$ is the Pauli matrix, the sign "$+$" corresponds to $\widetilde{V}^{(tx)}$ and the sign "$-$" to $\widetilde{V}^{(ty)}$.
$\xi^{(tx)}_{\varphi,\varphi'} \approx 1-(k_p^2a^2/8)(\cos\varphi + \cos\varphi')^2 $ and
$\xi^{(ty)}_{\varphi,\varphi'} \approx 1-(k_p^2 a^2/8)(\sin\varphi + \sin\varphi')^2 $.  $\xi^{(tx)}_{\varphi,\varphi'}$ and $\xi^{(ty)}_{\varphi,\varphi'}$
indicate that scattering on asymmetric impurities exhibits some angle dependence, but it is weak when the magnitude of the wavevector is small. However, what is crucial is that $\sigma_z$ does not commute with the orbital momentum operator.

To understand the effect of scattering on asymmetric impurities, characterized by the matrix elements $\widetilde{V}^{(tx)}$, we consider scattering from angle $\varphi$ to angle $\varphi'$ in the leading order with respect to the small parameter $\alpha$ (i.e. neglecting all the terms containing $\alpha$). Initially the distribution function is described by the diagonal terms $G_x(\varphi)$ and $G_y(\varphi)$, as well as the complex off-diagonal term $G_{xy}(\varphi)$. Its evolution is governed by the following equations:
\begin{subequations}
\begin{multline}
\frac{\partial G_x(\varphi)}{\partial t} = -\frac{2}{\tau_{tx}} G_x(\varphi), \quad \frac{\partial G_y(\varphi)}{\partial t} = 0
\\
\frac{\partial G_{xy}(\varphi)}{\partial t} = -\frac{1}{\tau_{tx}} G_{xy}(\varphi)
\end{multline}
\begin{multline}
\frac{\partial G_x(\varphi')}{\partial t} = \frac{2}{\tau_{tx}} G_x(\varphi), \\ \frac{\partial G_y(\varphi')}{\partial t} = \frac{\partial G_{xy}(\varphi')}{\partial t}= 0
\end{multline}
\end{subequations}
Here $\tau_{tx}^{-1} = 2a^4n_I\gamma_p(u_t \xi_{\varphi\varphi'}^{(tx)})^2/\hbar$. These equations show that, in the zeroth approximation with respect ot $\alpha$, where impurities scatter only $px$-states, there is an out-scattering term for the off-diagonal part of the distribution function, but no in-scattering term. Physically, this implies that the orbital momentum is completely lost in a single scattering event, corresponding to the Elliot-Yafet mechanism of relaxation.

Noteworthy that in the context of spin relaxation via the Elliot-Yafet mechanism, only a small fraction of spin polarization is typically lost in a single scattering event. This fraction is controlled by the spin-orbit coupling in the electron wavefunction and the impurity potential. In contrast,for orbital momentum, the spin-orbit interaction is replaced by the asymmetry of the impurity potential, which does not have a universal reason to be small and, therefore,  can lead to a very rapid orbital momentum relaxation.

While the Dyakonov-Perel mechanism can sometimes be suppressed by disorder, the Elliot-Yafet mechanism invariably becomes more pronounced as disorder increases. In this case,
the time dependence of the orbital momentum is described by the expression
\begin{equation}\label{map-EY}
L_z(t) = L_z(0)e^{-t/\tau_{Lz}} J_0(\omega_0 t)
\end{equation}
Here $\tau_{Lz} \propto \tau_{out}$ is the orbital momentum relaxation time, and $J_0$ is the Bessel function. Note that in the limit $\tau_{out} \rightarrow \infty$ the solution of Eq.~(\ref{map}) is also reduced to Eq.~(\ref{map-EY}).

\section{Skew-scattering orbital Hall effect}
\label{sect:skew}

In this section, we introduce a second toy model designed to demonstrate how scattering from impurities can lead to the skew-scattering mechanism of orbital Hall effect, while remaining as simple as possible. To achieve this, we consider the case where $t_{sp}=0$ which removes the orbital texture from the pristine material. However, we assume some degree of $s-p$ mixing in impurities represented by the contribution $V^{(sp)}$ to their energy (see Eq.~(\ref{Vsp})). Specifically, we assume that most of the impurities are strongly asymmetric and provide rapid wavevector and orbital momentum relaxation, but some impurities have the potential $V^{(sym)} + V^{(sp)}$ with $u_{sp}\ll u_I$. This model ensures that all the orbital Hall effect is entirely due to the skew scattering mechanism.

Within the quasi-classical approximation, impurity-induced $s-p$ mixing can be important only when both $s$ and $p$ states contribute to the Fermi surface. Therefore, unlike in Sec.~\ref{sect:EY}, we consider two Fermi wavevectors: $k_s$ corresponding to $s$-states, and $k_p$ corresponding to both $p_x$ and $p_y$ states.

Despite our best efforts, we have found no evidence of skew scattering when we restrict ourselves to the Born approximation. This situation is similar to the conventional skew-scattering spin-Hall effect, which only appears in third-order perturbation theory with respect to the impurity potential \cite{SHrev}. Consequently, in this section we include the contribution of the term beyond the Born approximation in the scattering by "orbital texture" impurities, although we consider this contribution to be small. We divide the scattering operator into the two parts. The main part, which describes the scattering by the asymmetric impurities, is reduced  to a phenomenological relaxation time $\tau_0$. The second part of the scattering operator is much smaller but is responsible for the skew scattering orbital Hall effect. It represents the  scattering by the "orbital texture" impurities treated by the third-order perturbation theory.

Similar to Sec.~\ref{sect:EY}, we assume that $|t_{p,\sigma} - t_{p\pi}|$ is small and the distribution function is confined to the narrow region of $k$-space near the Fermi surfaces. This allows us to integrate over the absolute value of the wavevector. However, in this model, there are two $k$-integrated distribution functions: $G_s$ and $G_p$ corresponding to the Fermi surfaces of $s$ and $p$-states respectively.

\begin{equation}\label{Gs}
G_s = \int_{k_s - \Delta k}^{k_s+\Delta k}f_1({\bf k}) k dk
\end{equation}
where $\Delta k$ is some value that is sufficiently large to include all the relevant momenta but small enough not to mix $k_s$ and $k_p$. Only the diagonal term of the distribution function related to $s$-states exists near $k_s$, making $G_s$ effectively a scalar.

\begin{equation}\label{Gp}
G_p = \int_{k_p - \Delta k}^{k_p+\Delta k}f_1({\bf k}) k dk
\end{equation}
is the second integrated distribution function related to $p_x$ and $p_y$ states, including their coherent superposition. It corresponds to the distribution function $f_1({\bf k})$ with $|{\bf k}| \sim k_p$. $G_p$ is effectively a $2\times 2$ matrix.
The coherent superpositions of $s$ and $p$ states precess with a very high frequency $\sim (\varepsilon_s({\bf k}) - \varepsilon_p({\bf k}))/\hbar$, and are therefore neglected.

The Boltzmann equation for $G_s$ and $G_p$ in a static electric field is given by
\begin{equation}
\frac{i}{\hbar}[H,G_{s,p}] + \frac{G_{s,p}}{\tau_0} = \hat{I}_3 G + {\cal I}_{s,p}
\end{equation}

Here $\hat{I}_3$ represents the part of the scattering operator that describes skew scattering and ${\cal I}_{s,p}$ is the source of the non-equilibrium distribution functions due to the applied electric field:
\begin{equation}\label{calI}
{\cal I}_{s,p} = 2\pi \gamma_{s,p} e {\bf E} {\bf v}_{s,p}
\end{equation}
Here ${\bf v}_{s,p}$ is the velocity in $s$ and $p$ bands respectively, $\gamma_s = (4\pi t_{s}a^2)^{-1}$ is the density od states in the $s$-band. Note that with $t_{sp} = 0$, ${\cal I}_{s,p} $ does not contain non-diagonal terms and does not lead to the intrinsic orbital Hall effect.

We expand $G_{s,p}$ in a series, treating $\hat{I}_3$ as a small parameter. $G_{s,p}^{(0)} = \tau_0{\cal I}_{s,p}$ is the solution of the Boltzmann equation in the Born approximation. The term $G_{s,p}^{(1)}$, which describes the skew-scattering orbital Hall effect, can be found from the equation:
\begin{equation}\label{Gsp1}
\frac{i}{\hbar}[H,G_{s,p}^{(1)}] + \frac{G_{s,p}^{(1)}}{\tau_0} = (\hat{I}_3 G^{(0)})_{s,p}
\end{equation}

The action of the scattering operator $\hat{I}_3$ on $G^{(0)}$ is considered in Appendix \ref{app:G3}. Here, we present the part that describes the generation of the orbital momentum due to the skew scattering, which is proportional to the Pauli matrix $\sigma_y$ in the $p$-band space: $(\hat{I}_3 G^{(0)})_p = S_y \sigma_y $.
\begin{multline}\label{IG0}
S_y = -\frac{ 6 \pi^2 u_{sp}^2 u_0 a^8 n_{sp}}{\hbar^2}  \\
\times \sin\varphi \left( \gamma_s\gamma_p k_p^3 + (\gamma_s\gamma_p - \gamma_p^2) k_s^2k_p \right) eE\tau_0.
\end{multline}
Here $n_{sp}$ is the 2D concentration of the "orbital texture" impurities.

The solution of Eq.~(\ref{Gsp1}) yields
\begin{equation}\label{Gp1}
G_p^{(1)}(\varphi) = \frac{S_y \tau_{0}}{1 + \omega^2(\varphi)\tau_{0}^2} (\sigma_y - \omega(\varphi)\tau_{0} \sigma_x)
\end{equation}
Here $\omega(\varphi) = k_p^2a^2 (t_{p,\sigma} - t_{p\pi})\cos(2\varphi)/\hbar$.

\begin{figure}
  \centering
  \includegraphics[width=0.45\textwidth]{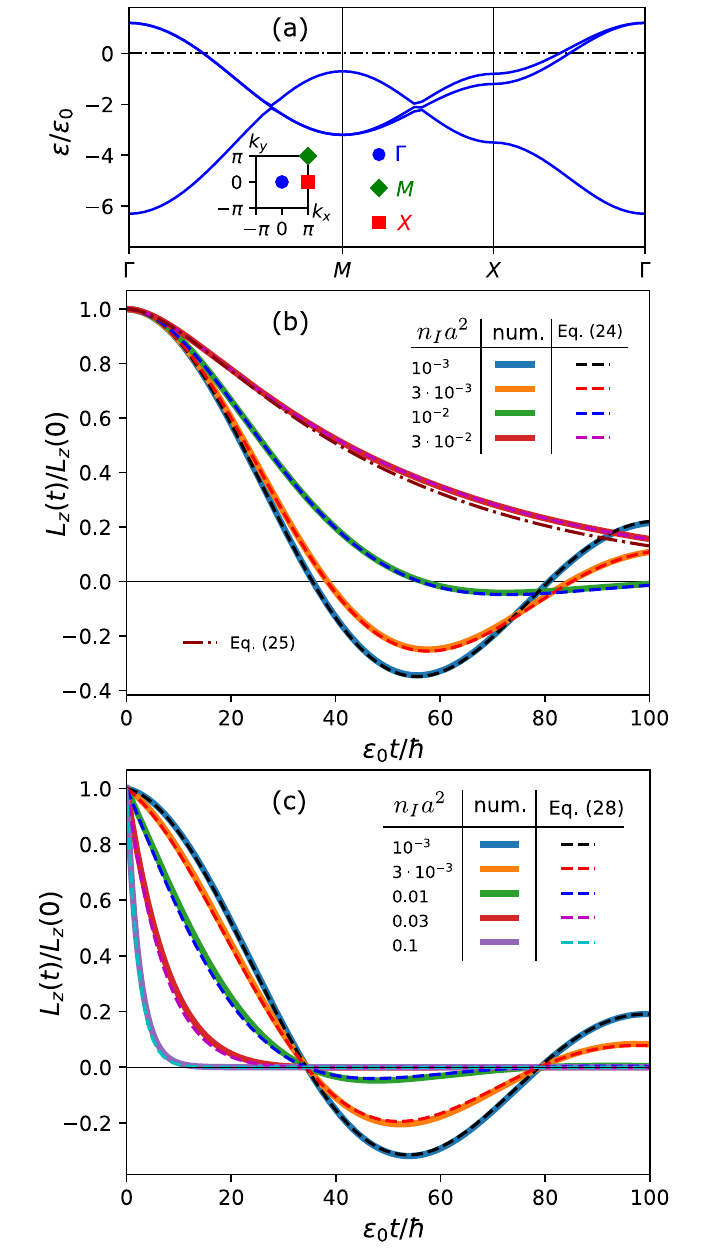}
  \caption{The comparison between the analytical description of the orbital momentum relaxation and the simulation results for "Set 1" parameters.
    (a) The band diagram corresponding to "Set 1", with the notations ($\Gamma$, $M$ and $X$ points) explained in the inset.
    (b) The Dyakonov-Perel orbital momentum relaxation due to scattering by symmetric impurities. The solid lines represent the numeric results for different impurity concentrations, as indicated in legend. The dashed lines show the results from solving Eq.~(\ref{map}) for the same impurity concentrations. The dash-dotted line represents  Eq.~(\ref{map-DP}) for $n_Ia^2 = 0.03$. (c) The Elliot-Yafet orbital momentum relaxation due to the scattering by asymmetric impurities. Solid lines represent the numerical results for the concentrations indicated in legend and dashed lines correspond to  Eq.~(\ref{map-EY}).}
  \label{fig:Rset1}
\end{figure}

The orbital momentum current density  $J_{OH}$ due to the skew scattering Hall effect can be expressed from $G_p^{(1)}$ as follows
\begin{equation}\label{Jskew1}
J_{OH} = \frac{\hbar}{(2\pi)^2} \int_0^{2\pi} v_p^{(y)} {\rm Tr} \left( G_p^{(1)}\sigma_y  \right) d\varphi
\end{equation}
Here $v_p^{(y)} = 2t_{p,\sigma}k_pa^2\sin(\varphi)/\hbar$ is the velocity in $y$-direction in the $p$-band. The integration yields
\begin{multline}\label{Jskew2new}
J_{OH} = - \frac{6\pi eE \tau_0^2 n_{sp} k_p^2a^{10} u_I u_{sp}^2t_{p,\sigma}}{\hbar^2 \sqrt{1+
\frac{k_p^4a^4}{\hbar^2} (t_{p,\sigma} - t_{p\pi})^2 \tau_0^2}} \\
 \times (k_p^2 \gamma_s\gamma_k  + k_s^2(\gamma_s\gamma_p-\gamma_p^2)).
\end{multline}

It is interesting to compare skew-scattering orbital Hall effect with the conventional skew-scattering spin-Hall effect. Usually, the skew-scattering spin current is inversely proportional to the impurity concentration,  making skew-scattering the dominant mechanism for the spin-Hall effect. Now let's consider the total impurity concentration, $n_I$, as a variable parameter, and assume that the fraction $n_{sp}/n_I$ remains constant as $n_I$ varies. The relaxation time $\tau_0$ is inversely proportional to  $n_I$ leading to the following dependence
\begin{equation}\label{Jskew3}
J_{OH} \propto \frac{\beta n_I^{-1}}{\sqrt{1 + \beta^2 n_I^{-2}}}
\end{equation}
where $\beta$ is a constant.

Eq.~(\ref{Jskew3}) shows that in contrast to the spin-Hall effect, skew scattering does not necessarily dominate over the intrinsic orbital Hall effect in cleaner materials with ${n_I \rightarrow 0}$.

\section{General case}
\label{sect:gen}

In this section, we discuss the numeric solution of Eq.~(\ref{kin1}).  Unlike the analytical approaches used in Sec.~\ref{sect:EY} and Sec.~\ref{sect:skew}, our numerical method imposes no restrictions on the model parameters. Moreover, the techniques presented here can be extended to more complex tight-binding models.

To solve Eq.~(\ref{kin1}) we decompose the scattering operator $\hat{I}$ into two components: the out-scattering operator $\hat{I}_{out}$ and in-scattering term $\hat{I}_{in}$. The distribution function $f_1$ is  expressed as a  series expansion: $f_1 = \sum_{m=0}^{\infty} f_1^{(m)}$, where the index $m$ represents the number of scattering events. The terms $f_1^{(m)}$ are related through a system of equations
\begin{subequations}\label{kinM}
\begin{equation}\label{kinM1}
\frac{\partial f_1^{(m)}}{\partial t}  +  \frac{i}{\hbar}\left[H, f_1^{(m)}\right] - \hat{I}_{out} f_1^{(m)}  = \hat{I}_{in} f_1^{(m-1)}
\end{equation}
\begin{equation}\label{kinM0}
\frac{\partial f_1^{(0)}}{\partial t}  + \frac{i}{\hbar}\left[H, f_1^{(0)}\right] - \hat{I}_{out} f_1^{(0)} = -\frac{e {\bf E}  }{\hbar} \frac{\partial f_0}{\partial {\bf k}}
\end{equation}
\end{subequations}

In this article, we focus on two scenarios: orbital momentum relaxation and orbital Hall effect under stationary conditions. For the former, we set ${\bf E} = 0$, while for the later we neglect the time derivative $\partial f_1^{(m)}/\partial t$.

For our simulation, the entire $k$-space is mapped onto an $N_k\times N_k$ square lattice, with $N_k=64$ unless stated otherwise. To compute the scattering operators, we replace the Dirac $\delta$-functions with the Gauss functions with a width $\sigma(\varepsilon)$. The energy-dependent  $\sigma$ is selected to ensure  that each lattice node has a sufficient number of "energy neigbors" (typically between 25 and 100), defined as other lattice sites with energies close enough that the broadened delta function is at least $1/e$ of its maximum value. We verified that the results are robust against variations in this selection.

\begin{figure}[t]
  \centering
  \includegraphics[width=0.5\textwidth]{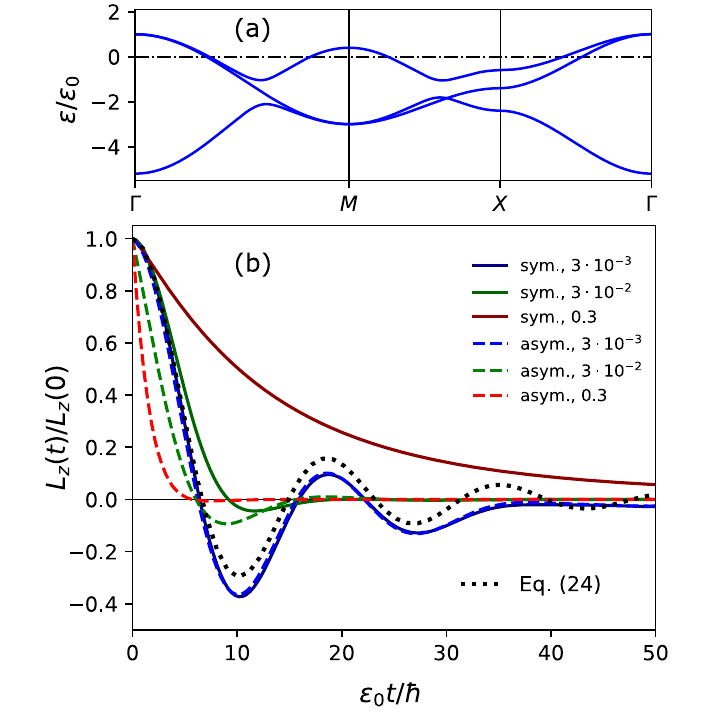}
  \caption{Orbital momentum relaxation for "Set 2" parameters. (a) The band diagram. (b) The results of simulation for different concentrations of symmetric or asymmetric as indicated in legend. Dotted line corresponds to the solution of Eq.~(\ref{map}).}
  \label{fig:Rset2}
\end{figure}

In the l.h.s. of Eqs.~(\ref{kinM}), there is no mixing of different wavevectors, which means that once $\hat{I}_{out}$ and $\hat{I}_{in} f_1^{(m-1)}$ are determined, the numerical solution of Eq.~(\ref{kinM1}) for $f_1^{(m)}$ becomes straightforward. The most computationally intensive step is calculating the in-scattering operator $\hat{I}_{in}$ acting on $f_1^{(m-1)}$. However, due to orbital momentum relaxation,
this calculation only needs to be performed a finite number of times. In some cases, such as when the Elliot-Yafet relaxation is strong, it is sufficient to compute only a few terms $f_1^{(m)}$ because the orbital momentum is lost after a small number of scattering events.

\begin{figure}[t]
  \centering
  \includegraphics[width=0.5\textwidth]{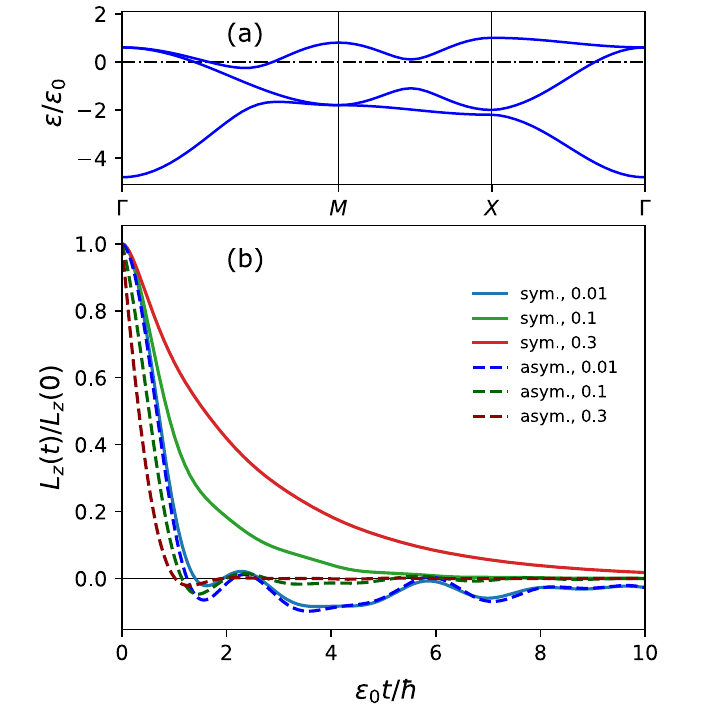}
  \caption{Orbital momentum relaxation for "Set 3" parameters. (a) The band diagram. (b) The results of simulation for different concentrations of symmetric or asymmetric as indicated in legend.}
  \label{fig:Rset3}
\end{figure}

We begin by presenting the results of our simulation of orbital momentum relaxation using the following set of parameters, referred to as "Set 1": $E_s = -3.5\varepsilon_0$, $E_p = -1.0\varepsilon_0$, $t_s = -0.7\varepsilon_0$, $t_{p,\sigma} = 0.6\varepsilon_0$, $t_{p,\pi}=0.55\varepsilon_0$ and $t_{sp} = 0.02\varepsilon_0$. Here $\varepsilon_0$ is an energy scale comparable to the bandwidth. The chemical potential is set to zero. The band structure corresponding to these parameters is shown in Fig.~\ref{fig:Rset1}(a).  "Set 1" satisfies the conditions for the applicability of  Eq.~(\ref{bolG}), with $\omega_0 = 0.07\varepsilon_0/\hbar$.

To study the relaxation we begin from a non-equilibrium distribution function $f_{in}$ that represents equilibrium in a hypothetical "orbital field": $B_L$ acting on the orbital momentum $L_z$. This distribution is described by the expression
\begin{equation}\label{fin}
f_{in} = \frac{1}{1+\exp\left(\frac{H'}{T}\right)}, \quad H' = H - B_L L_z
\end{equation}
In the simulations we used the parameters $B_L = 0.03\varepsilon_0/\hbar$ and $T=0.02\varepsilon_0$. The time-dependent quantum Boltzman equation is solved with this initial distribution function, and the resulting time-dependent orbital momentum is calculated as $\sum_{{\bf k}}\left({\rm Tr }L_zf_1({\bf k}, t)\right)/N_k^2$, where $L_z$ is described by Eq.~(\ref{Lz}).

The results of the simulation are presented in Fig.~\ref{fig:Rset1}(b,c). They demonstrate that the findings of Sec.~\ref{sect:EY} are in excellent agreement with the numerical calculations. Fig.~\ref{fig:Rset1}(b) illustrates the scenario where scattering is caused by symmetric impurities with $u_I= 2\varepsilon_0$ and different concentrations. The simulation results match the solution of Eq.~(\ref{map}). When scattering is strong and Dyakonov-Perel relaxation is suppressed, the numerical results also align with  Eq.~(\ref{map-DP}).  Fig.~\ref{fig:Rset1}(c) shows the results for scattering by asymmetric impurities. In this case,  half of the impurities are described by the potential $V^{(sym)} + V^{(tx)}$, and the other half by $V^{(sym)} + V^{(ty)}$. For both cases, the values $u_I = u_t = \varepsilon_0$ are used. The simulation results are in a very good agreement with Eq.~(\ref{map-EY}). In all analytical results (Eqs.~(\ref{map}, \ref{map-DP},\ref{map-EY}) we used the value $\tau_0 = 0.23\hbar/\varepsilon_0 n_Ia^2$.

To assess the robustness of the predictions made in Sec.~\ref{sect:EY} we introduce two additional sets of parameters. The second set ("Set 2") includes the values $E_s = -2.4\varepsilon_0$, $E_p = -1.0\varepsilon_0$, $t_s = -0.7\varepsilon_0$, $t_{p,\sigma} = 0.6\varepsilon_0$, $t_{p,\pi}=0.4\varepsilon_0$ and $t_{sp} = 0.2\varepsilon_0$. In this configuration, the difference between $t_{p,\sigma}$ and $t_{p,\pi}$ is still relatively small but larger than in "Set 1" and $t_{sp}$ is comparable to $t_{p,\sigma}-t_{p,\pi}$. Additionally, both $s$-band and $p$-bands contribute to the Fermi surface as shown in Fig.~\ref{fig:Rset2}(a). The third set ("Set 3") represents a more general scenario where the conditions of Sec.~\ref{sect:EY} are not met. It includes parameters $E_s = -2\varepsilon_0$, $E_p = -0.6\varepsilon_0$, $t_s = -0.7\varepsilon_0$, $t_{p,\sigma} = 0.7\varepsilon_0$, $t_{p,\pi}=-0.1\varepsilon_0$ and $t_{sp} = 0.3\varepsilon_0$. In this case, $t_{p,\sigma}$ and $t_{p,\pi}$ have opposite signs, while $t_{sp}$ exceeds the absolute value of $t_{p,\pi}$. The corresponding bands for these parameters are shown in Fig.~\ref{fig:Rset3}(a). We use the same values of $u_I$ and $u_t$ as in the previous scenario.

The simulation results for orbital momentum relaxation using "Set 2" are depicted in Fig.~\ref{fig:Rset2}(b). Although the agreement between Eq.~(\ref{map}) and the simulation is not as strong as in the previous case, the results from the Sec.~\ref{sect:EY} still qualitatively describe the numerical results. We observe a decaying oscillation of orbital momentum at low impurity concentrations. At high concentrations, symmetric impurities suppress the Dyakonov-Perel relaxation, whereas asymmetric impurities enhance relaxation through the Elliot-Yafet mechanism.  Fig.~\ref{fig:Rset3}(b) illustrates the simulated relaxation of orbital momentum for "Set 3". In this scenario, the oscillation of $L_z(t)$  displays interference of multiple frequencies. Despite this complexity, the qualitative effect of symmetric and asymmetric impurities remain consistent with those observed in other cases. However, due to the strong dephasing a high impurity concentration is necessary to significantly impact relaxation.

Next, we discuss the influence of different impurity types on the orbital Hall effect. We solve the static quantum Boltzmann equation under a small applied electric field in $x$-direction and calculate the orbital current by integrating the resulting distribution function with the operator $(L_z v_y + v_y L_z)/2$. Here $v_y = (\partial H/\partial  k_y)/\hbar$, note that it does not commute with $L_z$.

\begin{figure}[t]
  \centering
  \includegraphics[width=0.5\textwidth]{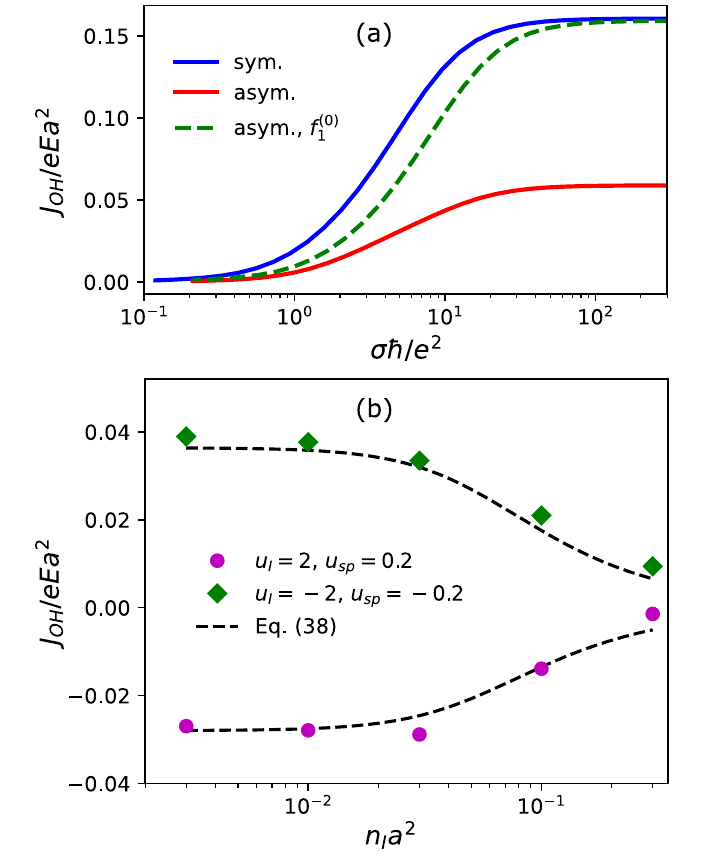}
  \caption{(a) The intrinsic orbital Hall effect calculated with symmetric and asymmetric impurities in the Born approximation as a function of conductivity. "Set 2" parameters are used for the calculation. The green dashed line corresponds to the orbital Hall effect calculated with asymmetric impurities without in-scattering. (b) The skew scattering orbital Hall calculated with the modified "Set 2" parameters with $t_{sp} = 0$. The is calculated with third-order perturbation theory with the orbital texture impurity potential $V^{(sym)} + V^{(sp)}$. Two sets of values for $u_I$ and $u_{sp}$ are considered as indicated in legend. The dashed line correspond to Eq.~(\ref{Jskew3}).  }
  \label{fig:oh}
\end{figure}

Fig.~\ref{fig:oh}(a) presents the simulation results of the intrinsic orbital Hall effect, calculated in the Born approximation. We focus on  comparing the effects of symmetric and asymmetric impurities using the same parameters as in Fig.~\ref{fig:Rset2}. To compare the different impurity models, we plot the orbital Hall effect as a function of conductivity. While the impurity type's impact on the intrinsic orbital Hall effect is less pronounced than its impact on relaxation, it is still significant. Asymmetric impurities result in an intrinsic orbital Hall effect that is approximately $2.7$ times weaker than in the case of symmetric impurities.

Interestingly, this impact persists even at very high conductivities when the impurity concentration is low. We interpret this as follows: The intrinsic orbital Hall effect
is a process that converts non-equilibrium  electron flow into orbital current due to the orbital texture. In the absence of in-scattering and with weak out-scattering, the non-equilibrium distribution function reduces to $f_1^{(0)}$, and the orbital current generated by the electric field is controlled solely by the orbital texture and oscillation frequencies of the orbital momentum. However, in-scattering can modify this outcome. Scattering by symmetric impurities does not produces additional flow, allowing terms $f_1^{(m)}$ with $m\ge 1$ to be safely ignored. In contrast, asymmetric impurities tend to scatter electrons backward, and their corresponding matrix elements, $V^{(tx)}$ and $V^{(ty)}$, do not commute with the Hamiltonian. It leads to in-scattered electrons being  in a quantum superposition of $p_x$ and $p_y$ bands, adding extra terms to the orbital current that partially compensate for the term related to $f_1^{(0)}$. To support this interpretation, we calculate orbital currents for asymmetric impurities without considering the in-scattering (i.e., without $m \ge 1$ terms in $f_1$). The results are shown with a dashed green line in  Fig.~\ref{fig:oh}(a). At high conductivities, they converge with the orbital current calculated for symmetric impurities.

Fig.~\ref{fig:oh}(b) presents the simulation results for the skew-scattering orbital Hall effect. In this simulation, we used a modified version of "Set 2" parameters with $t_{sp}=0$ to exclude the intrinsic orbital Hall effect from the results. We used $N_k=128$ for this simulation because, with $t_{sp}=0$, the orbital Hall effect has a significant contribution from small regions in $k$-space, necessitating a fine mesh in this space for accurate representation. This is particularly important for low impurity concentrations and long out-scattering times. The conventional skew-scattering spin Hall effect changes sign when the sign of the impurity potential is reversed. To investigate this, we calculated the orbital currents for the two cases of  orbital texture impurities with potentials $V^{(sym)} + V^{(sp)}$. The first case uses $u_I=2\varepsilon_0$ and $u_{sp}=0.2\varepsilon_0$, while the second case uses $u_I=-2\varepsilon_0$ and $u_{sp}=-0.2\varepsilon_0$. Scattering was calculated using  third-order perturbation theory.

Our results indicate that the skew-scattering orbital Hall effect exists and change sign when the impurity potential is reversed. Its dependence on the impurity concentration is qualitatively consistent with Eq.~(\ref{Jskew3}). However, distinguishing the skew-scattering orbital Hall effect from the intrinsic one in experiments would be challenging because, unlike the spin case, both mechansims exhibit a similar dependence on disorder.

\section{Discussion}
\label{sect:dis}

Our results reveal significant differences between the relaxation of spin and orbital momentum. Spin would be conserved in a solid state without the spin-orbit interaction. In materials with inversion symmetry,
the only mechanism of spin relaxation is the Elliot-Yafet mechansim, which arises from the interplay of the spin-orbit interaction and scattering. This mechanism results in a small probability of spin flip when an electron scatters off an impurity or phonon. When  inversion symmetry is broken, the Dyakonov-Perel mechanism comes into play, characterized by spin precession in an effective magnetic field that depends on the electron wavevector. However, these fields are also related to spin-orbit interaction and are usually small. As a result, in most cases, an electron undergoes many scattering events before its spin polarization is lost, making spin relaxation much slower than momentum relaxation. This condition is crucial for using the drift-diffusion equation for spin, a vital tool in the theory of spin transport~\cite{Dyak,RashbaSV,bouToka}.

Our study demonstrates, in line with the recent article~\cite{orbDP}, that the Dyakonov-Perel mechanism for orbital momentum relaxation exists even in materials with inversion symmetry because orbital momentum is always associated with coherent quantum superpositions of electron states across different bands. The corresponding effective fields are related not to the spin-orbit interaction but to the energy difference between bands, which can easily be large. Although Dyakonov-Perel orbital momentum relaxation can be suppressed by symmetric impurities, similar to the spin case, this requires that momentum relaxation be faster than orbital momentum dephasing. Therefore, high impurity concentrations are sometimes required to suppress Dyakonov-Perel relaxation of orbital momentum.

Moreover, the property of impurities that induces Elliot-Yafet relaxation of orbital momentum is asymmetry, unlike the spin-orbit interaction relevant for spin. This is not a relativistic effect, and there is no general requirement for it to be small. Consequently, sufficiently asymmetric impurities can cause orbital momentum relaxation in a single  scattering event. Interestingly, \textit{ab initio} calculations of orbital momentum relaxation for several metals, considering electron scattering on phonons, have shown in to be fast~\cite{DFTrel}. Although phonons are not explicitly considered in this article, the potential of electron-phonon interaction for a phonon with a finite wavevector lacks spherical symmetry and should contribute to at least some degree of Elliot-Yafet relaxation.

It appears that while the orbital momentum is much easier to polarize by current because it is not tied to spin-orbit interaction, it is also much easier to relax. Often, its relaxation would be comparable to or even faster than the relaxation of linear momentum. It makes the application of the drift-diffusion equations to orbital transport problematic. Although these equation are sometimes applied to orbital  momentum \cite{OHALL-Sala,Okin22}, such treatments are likely valid only under certain conditions, such as when Dyakonov-Perel relaxation is suppressed by a large number of symmetric impurities. Another consequence of rapid orbital momentum relaxation is its non-universality, meaning that the initial distribution of orbital momentum over wavevectors can influence the evolution of $L_z(t)$. In this article, we consider the system's initial state to correspond to the "orbital field" $B_L$. However, we expect that other initial states can lead to somewhat different dependencies $L_z(t)$.

All impurity models considered in this article preserve inversion symmetry. We have found that impurities breaking this symmetry can lead to the orbital Edelstein effect, i.e., the generation of an average orbital polarization by an applied current. However, this effect vanishes when multiple types of asymmetric impurities are introduced in such a way that inversion symmetry re-emerges after the averaging over impurity positions. On the other hand, Elliot-Yafet relaxation exists even in the case when an equal number of the impurities described by Eqs.~(\ref{Vpx}) and (\ref{Vpy}) are included in the simmulation (as was the case in all ours simulations with asymmetric impurities). For this mechanism of relaxation it is crucial that a single impurity breaks axial symmetry, regardless of the macroscopic symmetry of the averaged system.

In conclusion, we have developed a quasiclassical method based on the quantum Boltzmann equation to calculate the orbital momentum transport, incorporating scattering with a tight-binding impurity model.
Our findings demonstrate that the microscopic properties of impurities, particularly their symmetry, play a critical role in orbital momentum relaxation. When the impurity potential has axial symmetry,  relaxation occurs solely through the Dyakonov-Perel mechanism, which can be suppressed by increasing impurity concentration.  However, asymmetric impurities give rise to the very strong Elliot-Yafet relaxation, which is enhanced as impurity concentration increases.

We also report a significant influence of microscopic impurity properties on the orbital Hall effect. Even when scattering is considered within the Born approximation and only the intrinsic orbital Hall effect is taken into account, the specifics  of linear momentum relaxation kinetics affect the resulting orbital current.
Beyond the Born approximation, impurities that alter the orbital texture can induce a skew-scattering mechanism of the orbital Hall effect. Like its spin counterpart, this effect reverses sign when the impurity potential is inverted. However, unlike in the spin case, the skew-scattering mechanism does not necessarily dominate in nearly perfect crystals, making it more challenging to distinguishing from the intrinsic mechanism.

\appendix

\section{scattering operator}
\label{app:I3}

In this appendix, we derive the scattering operator $\hat{I}$ using the quantum mechanical perturbation theory. The quantum Boltzmann equation operates with the distribution
function, which is, in fact,  a density matrix in the Shrodinger representation. Within the quasiclassical approximation for a uniform system, only its matrix elements that are diagonal in
wavevector are considered, though these elements can still can be off-diagonal with respect to the band index. Here, we combine the wavevector and the band index into a single state index $a,b,c,...$ and introduce
\begin{equation}\label{fab}
f_{\alpha,\beta}({\bf k}) = f_{ab},
\end{equation}
where $a=(k,\alpha)$ and $b=(k,\beta)$. However, unlike $f_{\alpha,\beta}({\bf k})$, the notation $f_{ab}$ can also represent the off-diagonal in wavevector density matrix elements. Besides the notation $f_{ab}$ we use the density matrix $\rho_{ab}$ in the interaction representation.
\begin{equation}\label{rhoab}
\rho_{ab} = f_{ab}e^{i\omega_{ab}t}, \quad \omega_{ab} = \frac{\varepsilon_a - \varepsilon_b}{\hbar}
\end{equation}
Here $\varepsilon_a = \varepsilon_{\alpha}({\bf k})$ is the energy of electron in the state $a$ in the pristine material.

The perturbation series for $\rho_{ab}$ read
\begin{equation}\label{dr}
\frac{d \rho^{(n+1)}_{ac}}{dt} = -\frac{i}{\hbar} \left(V_{ab}'e^{i\omega_{ab}t}\rho^{(n)}_{bc}  - \rho^{(n)}_{ab} V_{bc}'e^{i\omega_{bc}t}  \right)
\end{equation}
Here, $V_{ab}'$ is the matrix element of the perturbation part of the Hamiltonian that accounts for impurities. The upper index $n$ in $\rho_{ab}^{(n)}$ shows the order of perturbation theory.
$\rho_{ab}^{(0)}$ corresponds the slowly varying distribution function $f_{ab}$, while the higher-order terms, starting from $\rho_{ab}^{(1)}$, contain rapid oscillations but are essential for deriving the scattering. We assume summation over repeating indices.

The formal integration of Eq.~(\ref{dr}) yields
\begin{equation}\label{r1}
\rho^{(1)}_{ac} = -\frac{1}{\hbar} \frac{ (V_{ab}' f_{bc} - f_{ab} V_{bc}'   ) e^{i\omega_{ac}t}}{\omega_{ac} - i\delta}
\end{equation}
Here $\delta \rightarrow +0$ and we have neglected the dependence of $f_{ab}$ and $f_{bc}$ on time.

The conventional derivation of the scattering operator in the Born approximation corresponds to the substitution of Eq.~(\ref{r1}) to Eq.~(\ref{dr}) and writing the expression for $f_{ab}^{(2)}$
\begin{multline}\label{df2}
\frac{d f^{(2)}_{ad}}{dt} = \frac{i}{\hbar^2}N_I \frac{V_{ab} (V_{bc} f_{cd} - f_{bc}V_{cd}  )}{\omega_{bd} - i\delta } \\
- \frac{i}{\hbar^2}N_I\frac{(V_{ab} f_{bc} - f_{ab}V_{bc}   )V_{cd}}{\omega_{ac} - i\delta}
\end{multline}
Here $N_I$ is the total number of impurities in the system and we took into account the following. The perturbation of the Hamiltonian $V'$ is composed of the contributions of the different impurities.
\begin{equation}\label{Vtild}
V_{ab}' = \sum_m e^{i({\bf k}' - \bf{k}) {\bf R}_m} V_{\alpha\beta}({\bf k},{\bf k'})
\end{equation}
Here, the index $m$ enumerates the impurities, and ${\bf R}_m$ denotes the position of each impurity. The description of the scattering involves averaging over these positions, leading to the substitution $V_{ab}'V_{ba}' \rightarrow N_I V_{ab}V_{ba}$, where $N_I = n_I{\cal S}/a^2$.

\begin{widetext}

To account for the skew scattering mechanism of the orbital Hall effect, however, we should repeat this procedure. This involves integrating the equation for $\rho_{ab}^{(2)}$ and obtaining its explicit expression.
\begin{equation}\label{rho2}
\rho^{(2)}_{ad} = \frac{e^{i\omega_{ad}t}}{\hbar^2 (\omega_{ad} - i\delta)} \left(    \frac{V_{ab}'
 ( V_{bc}'  f_{cd} - f_{bc} V_{cd}'  )}{\omega_{bd} - i\delta } - \frac{( V_{ab}'  f_{bc} - f_{ab} V_{bc}'    )V_{cd}' }{\omega_{ac} - i\delta}    \right)
\end{equation}
Then after its substitution to Eq.~(\ref{dr}) we get
\begin{multline}\label{df3}
\frac{d f^{(3)}_{ae}}{dt} = -\frac{i}{\hbar^3} N_I \left[
\frac{V_{ab}V_{bc}V_{cd} f_{de}}{(\omega_{be}-i\delta)(\omega_{ce}-i\delta)} -
\frac{V_{ab}V_{bc}f_{cd} V_{de}}{(\omega_{be}-i\delta)(\omega_{ce}-i\delta)} -
\frac{V_{ab}V_{bc}f_{cd} V_{de}}{(\omega_{be}-i\delta)(\omega_{bd}-i\delta)} +
\frac{V_{ab}f_{bc}V_{cd} V_{de}}{(\omega_{be}-i\delta)(\omega_{bd}-i\delta)}  \right.
\\    \left.
- \frac{V_{ab}V_{bc}f_{cd} V_{de}}{(\omega_{ad}-i\delta)(\omega_{bd}-i\delta)} +
\frac{V_{ab}f_{bc}V_{cd} V_{de}}{(\omega_{ad}-i\delta)(\omega_{bd}-i\delta)} +
\frac{V_{ab}f_{bc}V_{cd} V_{de}}{(\omega_{ad}-i\delta)(\omega_{ac}-i\delta)} -
\frac{f_{ab}V_{bc}V_{cd} V_{de}}{(\omega_{ad}-i\delta)(\omega_{ac}-i\delta)} \right]
\end{multline}

To use Eqs.~(\ref{df2}) and (\ref{df3}) as a scattering operator in Eq.~(\ref{kin1}), we apply the quasi-classical approximation. The denominators in Eqs.~(\ref{df2}) and (\ref{df3}) can be expressed as
$(\omega_{bd} - i\delta)^{-1} = {\rm p.v.}\, \omega_{bd}^{-1} + i\pi \delta(\omega_{bd})$. Here, the term with Dirac delta-function $i\pi \delta(\omega_{bd})$ corresponds to the transitions between states and the principal value ${\rm p.v.}\, \omega_{bd}^{-1}$ represents the impurity-induced correction to the electron states. The latter term is neglected under the quasi-classical approximation, implying that the electron states are weakly modified by disorder. Additionally, we neglect all the terms in l.h.s. of Eqs.~(\ref{df2}) and (\ref{df3}) that are not diagonal in wavevector. Under the Born approximation, it leads to Eq.~(\ref{Scat1}). Within the framework of third-order perturbation theory, we obtain the following.
\begin{multline}\label{df3k}
\frac{d f({\bf k})}{dt} = \frac{i\pi^2N_I}{\hbar} \sum_{{\bf k}',{\bf k}'',\eta,\lambda,\mu} \Bigl\{\delta(\varepsilon_\eta({\bf k}) - \varepsilon_\lambda({\bf k}'))\delta(\varepsilon_\eta({\bf k})-\varepsilon_\nu({\bf k}'')) \bigl[
\tilde{\delta}_{\eta} V({\bf k},{\bf k}')f({\bf k}')\tilde{\delta}_\lambda V({\bf k}',{\bf k}'')\tilde{\delta}_\nu V({\bf k}'',{\bf k}) - h.c.
\bigr]  \\
+ \delta(\varepsilon_\lambda({\bf k}') - \varepsilon_\eta({\bf k}))\delta(\varepsilon_\lambda({\bf k}')-\varepsilon_\nu({\bf k}'')) \bigl[
V({\bf k},{\bf k}')\tilde{\delta}_\lambda f({\bf k}') V({\bf k}',{\bf k}'')\tilde{\delta}_\nu V({\bf k}'',{\bf k})\tilde{\delta}_\eta - h.c.
\bigr] \\
+  \delta(\varepsilon_\nu({\bf k}'') - \varepsilon_\eta({\bf k}))(\varepsilon_\nu({\bf k}'')-\varepsilon_\lambda({\bf k}')) \bigl[
\tilde{\delta}_\eta V({\bf k},{\bf k}') \tilde{\delta}_\lambda f({\bf k}') V({\bf k}',{\bf k}'') \tilde{\delta}_\nu V({\bf k}'',{\bf k}) - h.c.
\bigr] \\
+ \delta(\varepsilon_\eta({\bf k}) - \varepsilon_\lambda({\bf k}'))(\varepsilon_\eta({\bf k})-\varepsilon_\nu({\bf k}'')) \Bigl[
  V({\bf k},{\bf k}'')\tilde{\delta}_\nu V({\bf k}'',{\bf k}') \tilde{\delta}_\lambda V({\bf k}'{\bf k})f({\bf k})\tilde{\delta}_\eta - h.c. \Bigr]
\Bigr\}
\end{multline}
Here $f({\bf k})$ and $V({\bf k},{\bf k}')$ are $3\times 3$ matrices in the band space. $\tilde{\delta}_{\eta}$ is a $3\times 3$ matrix with one on the $\eta$-th position on the diagonal and all the other elements equal zero. $h.c.$ means the Hermitian conjugate.  The complete expression of the scattering operator is given by the sum of the expressions (\ref{Scat1}) and (\ref{df3k}), however, we consider Eq.~(\ref{df3k}) as a small correction to the scattering operator and include it only when it is necessary to compute the skew-scattering Hall effect.

\section{Scattering operator integration over $|k|$}
\label{app:G3}

In this appendix, we integrate Eq.~(\ref{df3k}) over the absolute value $k={|{\bf k}|}$ under the assumptions made in Sec.~\ref{sect:skew}. We are interested in the expression for
\begin{equation}\label{a2-1}
\frac{d G_p(\varphi)}{dt} = \int_{k_p-\Delta k}^{k_p+\Delta k} \frac{d f_1({\bf k})}{dt} kdk
\end{equation}
According to our assumptions only the terms of $G_p$ corresponding to $p$-bands can be non-zero.

Let us focus on the first term in the r.h.s. of Eq.~(\ref{df3k}) and its Hermitian conjugate. They have the two important contributions different by the absolute value of ${\bf k}'$. The contribution (I) corresponds to $k_p-\Delta k < |{\bf k}'|<k_p+\Delta k$ and describes the scattering from $p$-states to $p$-states with an intermediate $s$-state. The contribution (II) represents the scattering from $s$ states to $p$-states and implies $k_s-\Delta k < |{\bf k}'|<k_s+\Delta k$. The expression for the contribution (I) with the assumptions made in Sec.~\ref{sect:skew} reads
\begin{multline}\label{dG3-a}
\frac{d [G_{p}(\varphi)]_{\alpha\beta}}{dt} =  \frac{i \pi^2 n_{sp} {\cal S}^3 }{\hbar(2\pi)^4}\int d\varphi'd\varphi''
\int_{k_p-\Delta k}^{k_p + \Delta k} kdk \, k'dk' \int_{k_s -\Delta k}^{k_s + \Delta k}  \, k''dk''\Bigl\{
\delta(\varepsilon_\alpha({\bf k}) - \varepsilon_\zeta({\bf k}') )
\delta(\varepsilon_\alpha({\bf k}) - \varepsilon_\chi({\bf k}'') )  \\
\times \widetilde{V}_{\alpha\xi}^{(pp)}(\varphi\varphi') [f_1({\bf k}')]_{\xi\zeta} \widetilde{V}_{\xi\chi}^{(ps)}(\varphi'\varphi'') \widetilde{V}_{\chi\beta}^{(sp)}(\varphi''\varphi)
\\ -
\delta(\varepsilon_\beta({\bf k}) - \varepsilon_\zeta({\bf k}') )
\delta(\varepsilon_\beta({\bf k}) - \varepsilon_\xi({\bf k}'') )
 \widetilde{V}_{\alpha\xi}^{(ps)} (\varphi\varphi'') \widetilde{V}_{\xi\zeta}^{(sp)}(\varphi''\varphi') [f_1({\bf k}')]_{\zeta\chi}  \widetilde{V}_{\chi\beta}^{(pp)}(\varphi'\varphi) \Bigr\}
\end{multline}
Here $V_{\alpha\beta}^{(ps)}(\varphi\varphi')$ corresponds to the matrix element of the impurity potential between the states $({\bf k}\alpha)$ and $({\bf k}'\beta)$ where $\alpha$ is one of the $p$-states, $|{\bf k}|$ is close to $k_p$, $\beta$ corresponds to $s$-state and $|{\bf k}'|\approx k_s$. Within our approximation such a matrix element depends only on the angles $\varphi$ and $\varphi'$.

The integrals over the absolute values $k$, $k'$ and $k''$ can be taken using Eq.~(\ref{Gp}) and the expressions $\gamma_s = (2\pi)^{-1} \int_{k_s -\Delta k}^{k_s + \Delta k} \delta(\varepsilon_s({\bf k}) - \varepsilon_F) kdk$ and
$\gamma_p = (2\pi)^{-1} \int_{k_p -\Delta k}^{k_p +\Delta k} \delta(\varepsilon_{p}({\bf k}) - \varepsilon_F)kdk$. The integration yields
\begin{multline}\label{dG3-a-1}
\frac{d G_{p}(\varphi)}{dt} =  \frac{i  n_{sp} {\cal S}^3 \gamma_s\gamma_p }{4\hbar}\int d\varphi'd\varphi''
 \Bigl\{
    \widetilde{V}^{(pp)}(\varphi\varphi') G_p(\varphi')  \widetilde{V}^{(ps)}(\varphi'\varphi'') \widetilde{V}^{(sp)}(\varphi''\varphi)
\\ -
 \widetilde{V}^{(ps)} (\varphi\varphi'') \widetilde{V}^{(sp)}(\varphi''\varphi') G_p(\varphi')  \widetilde{V}^{(pp)}(\varphi'\varphi) \Bigr\}
\end{multline}

In Sec.~\ref{sect:skew} we are interested in the calculation of the $\hat{I}_3$ acting on $G_p^{(0)}(\varphi) = eE\tau_0\cos\varphi k_p \hat{1}/\hbar$ where $\varphi$ is the angle between electric field and the wavevector and $\hat{1}$ is the unitary $2\times 2$ matrix. We also consider that the impurity potential is described by the combination of the symmetric potential and the orbital texture potential $V = V^{(0)} + V^{(sp)}$ with $u_{sp} \ll u_I$.
In this case Eq.~(\ref{dG3-a-1}) is reduced to
\begin{multline}\label{dG3-a-2}
\frac{d G_{p}(\varphi)}{dt} =  \frac{ n_{sp} a^8  \gamma_s\gamma_p u_I u_{sp}^2 eE\tau_0 k_p}{\hbar^2}\sigma_y \\
\times \int d\varphi'd\varphi'' \cos\varphi'
 \bigl[ k_p^2 (\cos\varphi\sin\varphi' - \cos\varphi'\sin\varphi) + k_sk_p\cos\varphi'' (\sin\varphi'-\sin\varphi) + k_sk_p\sin\varphi''(\cos\varphi - \cos\varphi')
   \bigr]
\end{multline}
The integral over the angles $\varphi'$ and $\varphi''$ equals to $-2\pi^2k_p^2\sin(\varphi)$ leading to
\begin{equation}\label{dG3-a-2}
\frac{d G_{p}(\varphi)}{dt} =  -\frac{ 2\pi^2 n_{sp} a^8  \gamma_s\gamma_p u_I u_{sp}^2 eE\tau_0 k_p^3}{ \hbar^2}\sin(\varphi)\sigma_y
\end{equation}
The similar treatments of the other terms in Eq.~(\ref{df3k}) allows one to derive Eq.~(\ref{IG0}).

\end{widetext}

\end{document}